\documentclass[aps,prl,twocolumn,showpacs,preprintnumbers,
               amsmath,amssymb]{revtex4}
\usepackage{graphicx}
\newcommand{\nn}{\nonumber}
\newcommand{\as}{\alpha_{s}}

\newcommand{\eqn}[1]{(\ref{#1})}
\newcommand{\mev}{\mbox{\rm MeV}}
\newcommand{\gev}{\mbox{\rm GeV}}

\begin{document}

\preprint{HD-THEP-04-30}
\preprint{IFIC/04-43}
\preprint{FTUV/04-0803}
\preprint{CAFPE/34-04}
\preprint{UG-FT/164-04}
\preprint{MPP-2004-93}
\preprint{TUM-HEP-555/04}

\title{\boldmath $V_{us}$ and $m_s$ from hadronic $\tau$ decays\unboldmath}

\author{Elvira G{\'a}miz${}^{\,a}$}
\author{Matthias Jamin${}^{\,b}$}
\author{Antonio Pich${}^{\,c}$}
\author{Joaquim Prades${}^{\,d}$}
\author{Felix Schwab${}^{\,e,f}$}

\affiliation{${}^a$Department of Physics and Astronomy, University of Glasgow,
Glasgow G12 8QQ, UK,}
\affiliation{${}^b$Institut f\"ur Theoretische Physik, Universit\"at
Heidelberg, Phil.-Weg 16, D-69120 Heidelberg, Germany,}
\affiliation{${}^c$Departament de F\'{\i}sica Te\`orica, IFIC, Universitat de
Val\`encia -- CSIC, Apt. Correus 22085, E-46071 Val\`encia, Spain,}
\affiliation{${}^d$Centro Andaluz de F\'{\i}sica de las Part\'{\i}culas
Elementales (CAFPE) and Departamento de F\'{\i}sica Te\'orica y del Cosmos,
Universidad de Granada, Campus de Fuente Nueva, E-18002 Granada, Spain,}
\affiliation{${}^e$Physik Department, Technische Universit\"at M\"unchen,
D-85748 Garching, Germany,}
\affiliation{${}^f$Max-Planck-Institut f\"ur Physik --
Werner-Heisenberg-Institut, D-80805 M\"unchen, Germany.}

\date{August 3, 2004}

\begin{abstract}
Recent experimental results on hadronic $\tau$ decays into strange particles
by the OPAL collaboration are employed to determine $V_{us}$ and $m_s$ from
moments of the invariant mass distribution. Our results are
$V_{us}=0.2208\pm0.0034$ and $m_s(2\,\gev)=81\pm22\,\mev$. The error on
$V_{us}$ is dominated by experiment, and should be improvable in the future.
Nevertheless, already now our result is competitive to the standard extraction
of $V_{us}$ from $K_{e3}$ decays, and it is compatible with unitarity.
\end{abstract}

\pacs{12.15.Ff, 14.60.Fg, 11.55.Hx, 12.38.Lg}

\maketitle


\section{Introduction}

Already more than a decade ago it was realised that the hadronic decay of the
$\tau$ lepton could serve as an ideal system to study low-energy QCD under
rather clean conditions \cite{bnp:92}. In the following years, detailed
investigations of the $\tau$ hadronic width as well as invariant mass
distributions have served to determine the QCD coupling $\alpha_s$ to a
precision competitive with the current world average \cite{aleph:98,opal:99}.
The experimental separation of the Cabibbo-allowed decays and
Cabibbo-suppressed modes into strange particles opened a means to also
determine the mass of the strange quark
\cite{gjpps:03,cdghpp:01,dchpp:01,km:00,kkp:00,pp:99,aleph:99,ckp:98,pp:98},
one of the fundamental QCD parameters within the Standard Model.

These determinations suffer from large QCD corrections to the contributions of
scalar and pseudoscalar correlation functions \cite{bnp:92,pp:98,ck:93,mal:98a}
which are additionally amplified by the particular weight functions which
appear in the $\tau$ sum rule. A natural remedy to circumvent this problem is
to replace the QCD expressions of scalar and pseudoscalar correlators by
corresponding phenomenological hadronic parametrisations
\cite{gjpps:03,mk:01,km:00,pp:99,aleph:99}, which turn out to be more precise
than their QCD counterparts, since the by far dominant contribution stems from
the well known kaon pole.

Additional suppressed contributions to the pseudoscalar correlators come from
the pion pole as well as higher exited pseudoscalar states whose parameters
have recently been estimated \cite{mk:02}. The remaining strangeness-changing
scalar spectral function has been extracted from a study of S-wave $K\pi$
scattering \cite{jop:00,jop:01} in the framework of resonance chiral
perturbation theory \cite{egpr:89}. The resulting scalar spectral function was
also employed to directly determine $m_s$ from a purely scalar QCD sum rule
\cite{jop:02}.

Nevertheless, as was already realised in the first works on strange mass
determinations from the Cabibbo-suppressed $\tau$ decays, $m_s$ turns out
to depend sensitively on the element $V_{us}$ of the quark-mixing (CKM) matrix.
With the theoretical improvements in the $\tau$ sum rule mentioned above, in
fact $V_{us}$ represents one of the dominant uncertainties for $m_s$. Thus it
appears natural to actually determine $V_{us}$ with an input for $m_s$ as
obtained from other sources \cite{gjpps:03}.

Very recently, new results on the $\tau$ branching fractions into strange
particles have been presented by CLEO \cite{cleo:03} and OPAL \cite{opal:04}.
In addition, the OPAL collaboration also presented an update on the strange
spectral function, previously known only from ALEPH \cite{aleph:99}. Both,
CLEO and OPAL found ${\cal B}[\tau^-\!\to K^-\pi^+\pi^-\nu_\tau]$ to be
significantly higher than the corresponding ALEPH result. The important impact
of these improved findings on the determination of $V_{us}$ and $m_s$ will be
investigated below.

\section{Theoretical framework}

The main quantity of interest for the following analysis is the hadronic decay
rate of the $\tau$ lepton,
\begin{equation}
\label{RTauex}
R_\tau \equiv \frac{\Gamma[\tau^-\to{\rm hadrons}\,\nu_\tau(\gamma)]}
{\Gamma[\tau^-\to e^-\bar\nu_e\nu_\tau(\gamma)]} = R_{\tau,NS} + R_{\tau,S}\,,
\end{equation}
which experimentally can be decomposed into a component with net-strangeness
$R_{\tau,S}$, and the non-strange part $R_{\tau,NS}$. Additional information
can be inferred from the measured invariant mass distribution of the final
state hadrons. The corresponding moments
$R_\tau^{kl}$, defined by \cite{dp:92b}
\begin{equation}
\label{Rtaukl}
R_\tau^{kl} \equiv \!\int\limits_0^{M_\tau^2}\! ds\, \biggl( 1 -
\frac{s}{M_\tau^2} \biggr)^{\!k}\!\biggl(\frac{s}{M_\tau^2}\biggr)^{\!l}
\frac{d R_\tau}{ds} = R_{\tau,NS}^{kl} + R_{\tau,S}^{kl} \,,
\end{equation}
can be calculated in complete analogy to $R_\tau = R_\tau^{00}$. In the
framework of the operator product expansion (OPE), $R_\tau^{kl}$ can be
written as
\cite{bnp:92}:
\begin{eqnarray}
R_\tau^{kl} &\!=\!& 3\,S_{{\rm EW}}\biggl\{\,\Big(|V_{ud}|^2+|V_{us}|^2\Big)
\,\Big( 1 + \delta^{kl(0)} \Big) \nn \\
&\!+\!& \sum\limits_{D\geq2}\Big( |V_{ud}|^2\,\delta_{ud}^{kl(D)} +
|V_{us}|^2\,\delta_{us}^{kl(D)} \Big)\,\biggr\} \,.
\end{eqnarray}
The electroweak radiative correction $S_{{\rm EW}}=1.0201\pm 0.0003$
\cite{ms:88,bl:90,erl:02} has been pulled out explicitly, and $\delta^{kl(0)}$
denotes the purely perturbative dimension-zero contribution. The symbols
$\delta_{ij}^{kl(D)}$ stand for higher dimensional corrections in the OPE from
dimension $D\geq 2$ operators which contain implicit suppression factors
$1/M_\tau^D$ \cite{ck:93,pp:98,pp:99}.

The separate measurement of Cabibbo-allowed as well as Cabibbo-suppressed decay
widths of the $\tau$ lepton \cite{aleph:99,cleo:03,opal:04} allows one to pin
down the flavour SU(3)-breaking effects, dominantly induced by the strange
quark mass. Defining the differences
\begin{equation}
\label{delRtaukl}
\delta R_\tau^{kl} \equiv \frac{R_{\tau,NS}^{kl}}{|V_{ud}|^2} -
\frac{R_{\tau,S}^{kl}}{|V_{us}|^2} =
3\,S_{{\rm EW}}\sum\limits_{D\geq 2}\Big(\delta_{ud}^{kl(D)} -
\delta_{us}^{kl(D)}\Big) \,,
\end{equation}
many theoretical uncertainties drop out since these observables vanish in the
SU(3) limit.

\section{Determination of \boldmath{$V_{us}$}}

\begin{figure}[thb]
\begin{center}
\includegraphics[angle=270, width=8cm]{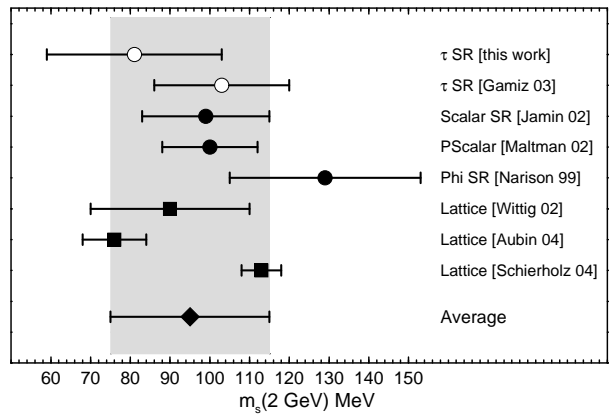}
\end{center}
\vspace{-5mm}
\caption{Summary of recent QCD sum rule \cite{gjpps:03,jop:02,mk:02,nar:99}
and lattice QCD \cite{wit:02,hpqcd:04,qcdsf:04} results for $m_s(2\,\gev)$.
For ref.~\cite{qcdsf:04} the given error is only statistical.\label{fig:ms}}
\end{figure}

Employing the SU(3)-breaking difference \eqn{delRtaukl}, as a first step, we 
intend to determine $V_{us}$. This approach requires a value for the strange
mass from other sources as an input so that we are in a position to calculate
$\delta R_\tau^{kl}$ from theory. In the following, we shall use the result
$m_s(2\,\gev)=95\pm 20\,\mev$, a value compatible with most recent
determinations of $m_s$ from QCD sum rules \cite{jop:02,mk:02,nar:99} and
lattice QCD \cite{wit:02,hpqcd:04,qcdsf:04}. The compilation of recent strange
mass determinations is displayed in figure~\ref{fig:ms}. For comparison,
in figure~\ref{fig:ms}, we also display $m_s$ as obtained from our previous
$\tau$ sum rule analysis \cite{gjpps:03} for the ALEPH data, as well as this
work analysing the OPAL data.

Since the sensitivity of $\delta R_\tau^{kl}$ to $V_{us}$ is strongest for
the (0,0) moment, where also the theoretical uncertainties are smallest, this
moment will be used for the determination of $V_{us}$. Inserting the above
strange mass value into the theoretical expression for
$\delta R_{\tau}$ \cite{gjpps:03}, one finds
\begin{equation}
\label{delR00th}
\delta R_{\tau,{\rm th}} \;=\; 0.218 \pm 0.026 \,,
\end{equation}
where the uncertainty dominantly results from a variation of $m_s$ within its
errors. Employing the above result in eq.~\eqn{delRtaukl}, together with the
experimental findings $R_{\tau,NS}=3.469\pm0.014$, $R_{\tau,S}=0.1677\pm0.0050$
\cite{opal:04}, as well as $|V_{ud}|=0.9738\pm0.0005$ \cite{pdg:04}, we then
obtain
\begin{equation}
\label{Vus}
|V_{us}| = 0.2208 \pm 0.0033_{\rm exp} \pm 0.0009_{\rm th} =
0.2208 \pm 0.0034\,.
\end{equation}
The first given error is the experimental uncertainty, dominantly due to
$R_{\tau,S}$, whereas the second error stems from the theoretical quantity
$\delta R_{\tau,{\rm th}}$. For the extraction of $V_{us}$, even though the
theoretical error on $\delta R_{\tau,{\rm th}}$ is 12\%, it represents only
a small correction compared to $R_{\tau,NS}/|V_{ud}|^2$ and thus its error is
suppressed. The theoretical uncertainty in $\delta R_{\tau,{\rm th}}$ will
only start to matter once the experimental error on $R_{\tau,S}$ is much
improved, possibly through analyses of the BABAR and BELLE $\tau$ data samples.

One further remark is in order. A sizeable fraction of the strange branching
ratio is due to the decay $\tau\to K\nu_\tau$, for which OPAL used the PDG fit
result ${\cal B}[\tau\to K\nu_\tau(\gamma)]=(0.686\pm0.023)\%$ \cite{pdg:04}.
However, this decay can be predicted employing its relation to the decay
$K\to\mu\nu_\mu(\gamma)$, which theoretically is known rather well
\cite{df:94,df:95}. Updating the numerics of refs.~\cite{df:94,df:95}, we
then obtain ${\cal B}[\tau\to K\nu_\tau(\gamma)]=(0.715\pm0.004)\%$, much more
precise than the experimental value. Adding this result to the remaining
strange branching fractions, one finds $R_{\tau,S}=0.1694\pm0.0049$, which
would lead to $|V_{us}|=0.2219 \pm 0.0034$.

\section{Strange quark mass}

Employing the above calculated value for $V_{us}$, we are now in a position
to determine the strange quark mass $m_s$ from the SU(3)-breaking difference
of eq.~\eqn{delRtaukl}. Experimentally, various $(k,l)$ moments have been
determined \cite{opal:04}. For low $k$, the higher-energy region of the
experimental spectrum, which is less well known, plays a larger role and thus
in this region the experimental uncertainties dominate the strange mass
determination, whereas for higher $k$ more emphasis is put on the lower-energy
region, and there the theoretical uncertainties dominate. At present, the most
reliable results for $m_s$ are obtained from the moments $(2,0)$ to $(4,0)$,
and we shall only discuss these here. 

\begin{table}[t]
\renewcommand{\arraystretch}{1.2}
\begin{center}
\begin{tabular}{ccrrr}
\hline
Parameter & Value & \quad(2,0)\quad & \quad(3,0)\quad & \quad(4,0)\quad \\
\hline
$m_s(M_\tau)$ & & 93.2 & 86.3 & 79.2 \\
\hline
$R_{\tau,{\rm NS}}^{kl}$ & \cite{opal:04} &
${}^{+5.1}_{-5.4}$ & ${}^{+3.6}_{-3.7}$ & ${}^{+2.8}_{-2.9}$ \\
$R_{\tau,{\rm S}}^{kl}$ & \cite{opal:04} &
${}^{-30.9}_{+23.3}$ & ${}^{-19.5}_{+15.8}$ & ${}^{-13.9}_{+11.6}$ \\
$|V_{us}|$ & $0.2208\pm0.0034$ &
${}^{+21.7}_{-29.8}$ & ${}^{+14.6}_{-18.7}$ & ${}^{+10.6}_{-13.0}$ \\
${\cal O}(\as^3)$ & ${}^{2\times{\cal O}(\as^3)}_{{\rm no}\;{\cal O}(\as^3)}$ &
${}^{-4.0}_{+4.6}$ & ${}^{-5.3}_{+6.5}$ & ${}^{-6.1}_{+7.8}$ \\
$\xi$ & ${}^{1.5}_{0.75}$ &
${}^{+2.7}_{+2.3}$ & ${}^{+4.7}_{-0.2}$ & ${}^{+6.3}_{-2.2}$ \\
$\as(M_\tau)$ & $0.334\pm0.022$ &
${}^{+0.7}_{+0.1}$ & ${}^{-0.7}_{+1.3}$ & ${}^{-1.6}_{+2.2}$ \\
$\langle\bar ss\rangle/\langle\bar uu\rangle$ & $0.8\pm0.2$ &
${}^{-8.7}_{+7.7}$ & ${}^{-9.9}_{+8.9}$ & ${}^{-10.6}_{\phantom{1}+9.6}$ \\
$f_K$ & $113\pm 2\,\mev$ &
${}^{-1.8}_{+1.7}$ & ${}^{-1.4}_{+1.4}$ & ${}^{-1.2}_{+1.1}$ \\
\hline
Total & &
${}^{+33.6}_{-44.3}$ & ${}^{+25.0}_{-29.5}$ & ${}^{+21.3}_{-23.0}$ \\
\hline
\end{tabular}
\end{center}
\caption{Central results for $m_s(M_\tau)$ extracted from the different
moments, as well as ranges for the main input parameters and resulting
uncertainties for $m_s$.\label{tab1}}
\end{table}

The analysis proceeds in complete analogy to our previous work \cite{gjpps:03}.
In table~\ref{tab1}, we show a detailed account of our results. The first row
displays the values of $m_s(M_\tau)$ obtained from the different moment sum
rules and central values for all input parameters. In the following rows, we
have listed those input parameters which dominantly contribute to the
uncertainty on $m_s$, the ranges for these parameters used in our analysis and
the resulting shift in $m_s$. Only those parameters have been included in the
list which at least for one moment yield a shift of $m_s$ larger than
$1\,\mev$. Finally, in the last row, we display the total error that results
from adding the individual uncertainties in quadrature.

Taking a weighted average of the strange mass values obtained
for the different moments we then find
\begin{eqnarray}
\label{ms}
m_s(M_\tau) &\!=\!& 84 \pm 23 \, \mev \,, \nn \\
\Rightarrow \quad m_s(2\,\gev) &\!=\!& 81 \pm 22 \, \mev \,,
\end{eqnarray}
where the uncertainty corresponds to that of the (4,0) moment. The dominant
theoretical uncertainties in the result of eq.~\eqn{ms} originate from higher
order perturbative corrections as well as the SU(3)-breaking ratio of the quark
condensates $\langle\bar ss\rangle/\langle\bar qq\rangle$ \cite{jam:02} which
arises in the dimension-4 contribution to eq.~\eqn{delRtaukl}. A detailed
discussion of all input parameters can be found in ref.~\cite{gjpps:03}.

In our previous analysis \cite{gjpps:03}, based on the ALEPH data
\cite{aleph:99}, it was observed that $m_s$ displayed a strong dependence on
the number of the moment $k$, decreasing with increasing $k$, and it was
speculated that this behaviour could be due to missing contributions in the
higher energy region of the spectrum \cite{jamin:03}. With the recent CLEO
and OPAL data \cite{cleo:03,opal:04}, finding a larger branching fraction of
the $K^-\pi^+\pi^-$ mode, the decrease of $m_s$ is now much reduced, although
still visible. This issue needs to be clarified further once even better data
are available.

\section{Simultaneous fit of \boldmath{$V_{us}$} and \boldmath{$m_s$}}

In principle, it is also possible to perform a simultaneous fit to $V_{us}$ and
$m_s$ from a certain set of $(k,l)$ moments. As soon as more precise data are
available, this will be the ultimate approach to determine $V_{us}$ and $m_s$
from hadronic $\tau$ decays. With the current uncertainties in the data and the
question about a monotonous $k$-dependence of $m_s$, a bias could be present
in the method. Furthermore, the correlations between different moments are
rather strong and also have to be included on the theory side.

Here, we shall restrict ourselves to a simplified approach where all
correlations are neglected. For the simultaneous fit of $V_{us}$ and $m_s$,
we employ the five $R_\tau^{kl}$ moments $(0,0)$ to $(4,0)$ which have also
been used in our previous analysis \cite{gjpps:03}. Performing this exercise,
for the central values we find:
\begin{equation}
\label{msfit}
|V_{us}| \;=\; 0.2196 \,, \qquad m_s(2\,\gev) = 76 \, \mev \,.
\end{equation}
The expected uncertainties on these results should be smaller than the
individual errors in eqs.~\eqn{Vus} and \eqn{ms}, but only slightly since the
correlations between different moments are rather strong.

The general trend of the fit result can be understood easily. $m_s$ from the
OPAL data turned out lower than our global average $m_s(2\,\gev)=95\pm20\,\mev$
considered in section~3. Thus, also the corresponding $\delta R_{\tau,{\rm th}}$
is lower, resulting in a slight reduction of $V_{us}$. Furthermore, the
moment-dependence of $m_s$ is reduced in the fit. Nevertheless, leaving a
detailed error analysis for a forthcoming publication \cite{gjpps:05}, at
present, we consider eqs.~\eqn{Vus} and \eqn{ms} as our central results.

\section{Conclusions}

Taking advantage of the strong sensitivity of the flavour-breaking $\tau$
sum rule on the CKM matrix element $V_{us}$, it is possible to determine
$V_{us}$ from hadronic $\tau$ decay data. This requires a value of the strange
quark mass as an input which can be obtained from other sources like QCD sum
rules or the lattice. The result for $V_{us}$ thus obtained is
\begin{equation}
|V_{us}| \;=\; 0.2208 \pm 0.0034 \,,
\end{equation}
where the error is largely dominated by the experimental uncertainty on
$R_{\tau,S}$, and thus should be improvable with the BABAR and BELLE $\tau$
data sets in the near future. Already now, our result is competitive with the
standard extraction of $V_{us}$ from $K_{e3}$ decays
\cite{lr:84,e865:03,ktev:04,ckm:03} and a new determination from $f_K/f_\pi$
as extracted from the lattice \cite{mar:04,milc:04}. The resulting deviation
from CKM unitarity then is
\begin{equation}
1 - |V_{ud}|^2 - |V_{us}|^2 - |V_{ub}|^2 = (2.9\pm 1.8)\cdot 10^{-3} \,,
\end{equation}
being consistent with unitarity at the $1.6\,\sigma$ level.

For the strange mass determination, we have used the three moments $(2,0)$
to $(4,0)$, with the result
\begin{equation}
m_s(2\,\gev) = 81 \pm 22 \, \mev \,.
\end{equation}
Our value for $m_s$ is on the low side of previous strange mass determinations,
but certainly compatible with them. It is also on the borderline of being
compatible with lower bounds on $m_s$ from sum rules
\cite{lrt:97,ynd:98,dn:98,ls:98,mk:02}.

Finally, we have performed a simultaneous fit of $V_{us}$ and $m_s$ to the five
moments $(0,0)$ to $(4,0)$. Our central values are completely compatible with
the central results of eqs.~\eqn{Vus} and \eqn{ms}. Anticipating a detailed
analysis of the correlations between different moments, these findings should
be considered as an indication of the prospects for the future when more
precise experimental data will become available.

\vspace{3mm}
\begin{acknowledgments}
It is a great pleasure to thank the Benasque Center for Science, where part of
this work has been performed, for support. This work has been supported in part
by the European Union RTN Network EURIDICE Grant No. HPRN-CT2002-00311 (A.P.
and J.P.), by MCYT (Spain) and FEDER (EU) Grants No. FPA2003-09298-C02-01
(J.P.), and FPA-2001-3031 (A.P. and M.J.), and by Junta de Andaluc\'{\i}a Grant
No. FQM-101 (J.P.). E.G. is indebted to the EU for a Marie-Curie-Fellowship.
M.J. would like to thank Ulrich~Uwer for helpful discussions.
\end{acknowledgments}


\end{document}